# Ultrabroadband THz/IR upconversion and photovoltaic response in semiconductor ratchet based upconverter

Peng Bai[1], Ning Yang[1], Weidong Chu[1, a)], Yueheng Zhang[2], Wenzhong Shen[2], Zhanglong Fu[3], Dixiang Shao[3], Kang Zhou[3], Zhiyong Tan[3], Hua Li[3], Juncheng Cao[3], Lianhe Li[4], Edmund Harold Linfield[4], Yan Xie[5], Ziran Zhao[5, b)]

1. *Institute of Applied Physics and Computational Mathematics, Beijing 100088, China*
2. *Key Laboratory of Artificial Structures and Quantum Control, School of Physics and Astronomy, Shanghai Jiao Tong University, Shanghai 200240, China*
3. *Key Laboratory of Terahertz Solid-State Technology, Shanghai Institute of Microsystem and Information Technology, Chinese Academy of Sciences, Shanghai 200050, China*
4. *School of Electronic and Electrical Engineering, University of Leeds, Leeds LS2 9JT, UK*
5. *Department of Engineering Physics, Tsinghua University, Beijing 100084, China*

E-mail addresses: a)chu_weidong@iapcm.ac.cn, b)zhaozr@mail.tsinghua.edu.cn

An ultrabroadband upconversion device is demonstrated by direct tandem integration of a p-type GaAs/Al$_x$Ga$_{1-x}$As ratchet photodetector (RP) with a GaAs double heterojunction LED (DH-LED) using the molecular beam epitaxy (MBE). An ultrabroadband photoresponse from terahertz (THz) to near infrared (NIR) region (4-200 THz) was realized that covers a much wider frequency range compared with the existing upconversion devices. Broadband IR/THz radiation from 1000 K blackbody is successfully upconverted into NIR photons which can be detected by commercial Si-based device. The normal incidence absorption of the RP simplifies the structure of the RP-LED device and make it more compact compared with the inter-subband transition based upconverters. In addition to the upconversion function, the proposed upconverter is also tested as photovoltaic detectors in the infrared region (15-200 THz) without an applied bias voltage due to the ratchet effect.

Upconversion process that convert the photons from lower to higher energies is promising for the applications ranging from high-efficiency solar cells to highly sensitive biological imaging and large format infrared (IR)/terahertz (THz) imaging [1-5]. A compact broadband upconverter is desirable to promote these applications. But until now the spectral conversion under broadband excitation (from near infrared (NIR) to THz) has been challenging [6]. The upconversion nanoparticles based on lanthanide compounds are widely used in efficiency improving of the photovoltaic and background-free bioimaging [1, 4, 7]. However, the absorption spectral coverage of these nanoparticles is mainly focus on the NIR range (<1 μm) and it's difficult to realize longer wavelength IR or THz absorption.

Another competitive up-conversion method is based on the sum-frequency generation in nonlinear optical crystal (periodically poled lithium niobate (PPLN)), which can convert THz radiation, mid infrared (MIR) photons or infrared telecom-band photons into NIR photons to be detected by a commercial Si diode or charge coupled device (CCD) [8-11]. This up-conversion technique requires an additional high-power laser pump source and depends on the nonlinear optical crystal, which cause complex optical setup and narrow spectral sensitivity. A 3D upconversion technique using the glow discharge detector (GDD) is a simple and inexpensive scheme and has good performance in millimeter wave (MMW)/THz imaging systems, but the performance of this approach in the MIR or NIR range needs further study [12]. A novel mechanism to transfer energy into higher-energy modes by exploiting the nonlinearities that couple phonons in $SrTiO_3$ [13] has also been reported. Whether one can control higher-frequency, mid-infrared modes via this nonlinear phonon up-conversion process also require further exploring.

Semiconductor-based upconversion is another promising up-conversion mechanism, in which IR or THz photons are first absorbed by a photodetector, and then the photo-generated carriers drive a LED to emit near-infrared light, thereby achieving up-conversion [5, 14]. The quantum well photodetector (QWP) based upconverter is a highly-developed upconverter [15-20]. Infrared thermal imaging devices as well as THz prototype imaging devices based on the QWP-LED have been reported [21, 17], in which the large format infrared imaging could be realized easily without the need to the integrate Si readout circuits on the chip and fully compatible with the semiconductor processing technology. However, the inter-subband transition (ISBT) mechanism in QWP limits the response spectral coverage of the device and requires the device to introduce an additional optical coupling structure to achieve normal incidence response [22]. An alternative approach is to use the GaAs homojunction interfacial workfunction internal photoemission (HIWIP) to fabricate the upconverter [23]. The HIWIP-LED upconverter allows normal incidence response and shows a broadband photoresponse, based on which a high-resolution THz imaging has been demonstrated, although the low activation energy of the HIWIP results in large dark current and requires extremely low temperature (~4 K) operation condition [24].

In this paper, an ultrabroadband semiconductor-based upconversion device is demonstrated, which is based on the integrated p-type $GaAs/Al_xGa_{1-x}As$ ratchet photodetector (RP) and GaAs double heterojunction LED (DH-LED). An ultrabroadband photoresponse from THz to NIR region (4-200 THz) covers a much wider frequency range compared with the other upconversion devices. The RP-LED permits normal incidence excitation thus bypassing the need for the input coupler required for ISBT-based upconverters. The upconverter also operates as a photovoltaic detector at zero bias with a response range from 15-200 THz owing to the ratchet effect.

The structure diagram of the RP-LED upconverter is shown in Fig.1(a), in which the upconversion process is indicated: the incoming THz or IR photons are absorbed in the ratchet detector and converted into photo-carriers. Then the photogenerated carriers are injected into the active region of the LED under the electric field to recombine and emit NIR photons. The band gap profile representation of upconversion mechanism of RP-LED is shown in Fig.1(b).

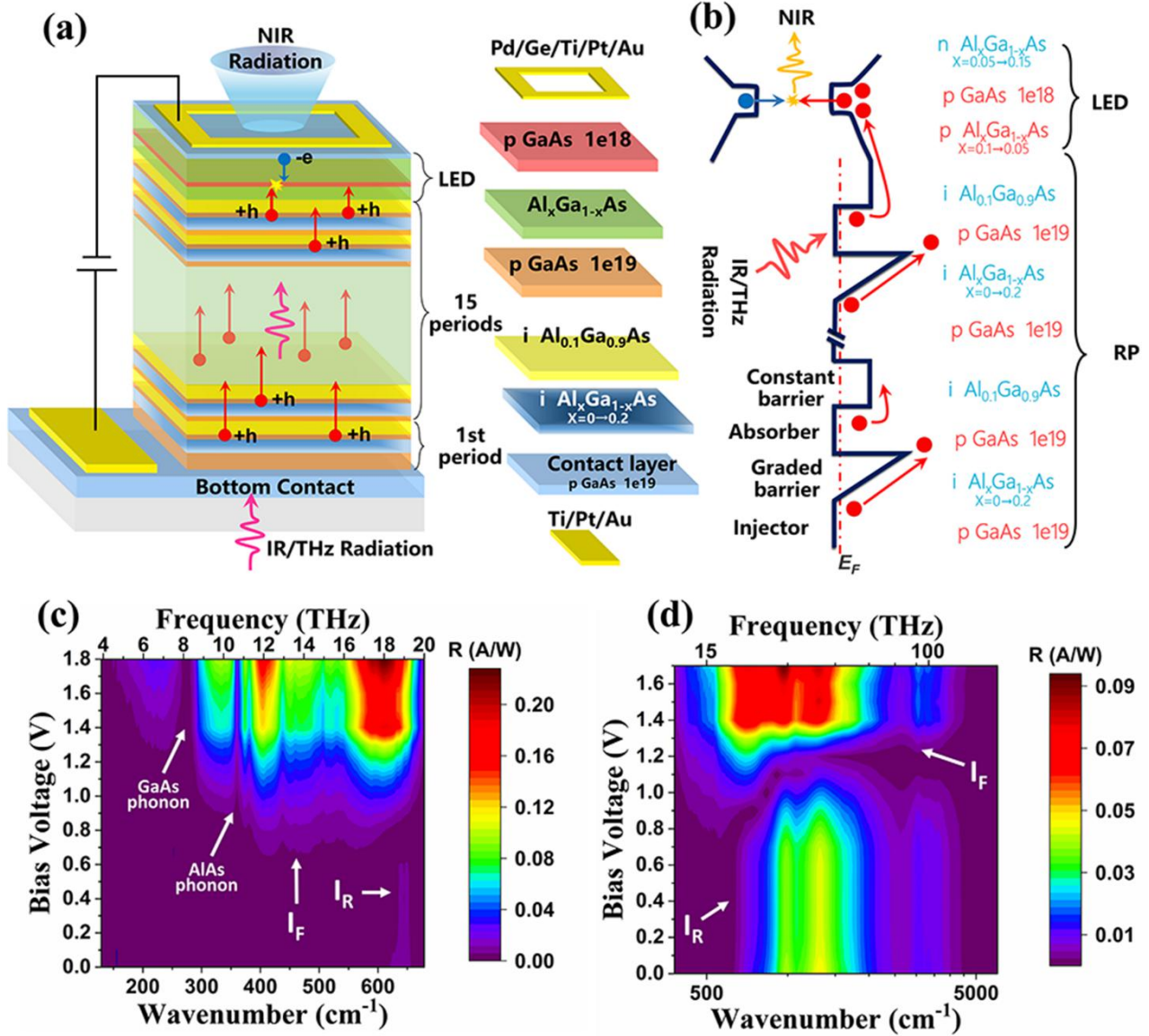

**Figure 1.** (a) Structure diagram of the RP-LED upconverter, (b) the band diagram and upconversion mechanism of RP-LED device, (c) the responsivities with different bias voltages at the frequency range of 4- 20 THz and (d) 10.5- 225 THz. The arrow indicated $I_F$ means that the photocurrent is consistent with the direction of the turn-on bias voltage of the LED (with the top contact being grounded) and the $I_R$ means that the photocurrent is opposite to the direction of the turn-on bias voltage of the LED.

The epitaxial growth of RP-LED structure was made on a semi-insulating (100) GaAs substrate. It began with a 20-nm-thick intrinsic buffer layer doped to a Be concentration of $1\times10^{19}$ cm$^{-3}$ followed by the RP in the form of its 500-nm-thick bottom p-GaAs contact layer continuing with the first $Al_xGa_{1-x}As$ ratchet barrier, first 20-nm p-GaAs absorber and first $Al_{0.1}Ga_{0.9}As$ constant barrier. It was followed by the multiple ratchet structure growth which consisted of a 15 repeat of p-GaAs / $Al_xGa_{1-x}As$/ p-GaAs/ $Al_{0.1}Ga_{0.9}As$ (20/80/20/80 nm) with x changing from 0 to 0.2. The growth continued with the LED constituents: a 40-nm-thick p-$Al_xGa_{1-x}As$ ($1\times10^{19}$ cm$^{-3}$) graded layer with x=0.2 at the beginning and decreasing to x=0.05 at the end followed by a 150-nm-thick p-GaAs layer, then a 40-nm-thick graded n-$Al_xGa_{1-x}As$ with x=0.05 at start and x=0.1 at the end. This was followed by an 80-nm-thick n-Al$_{0.1}$Ga$_{0.9}$As, and finally, a 400 nm n-GaAs top contact layer. All the p-GaAs in the active region doped to a Be concentration of 1×10$^{19}$ cm$^{-3}$ and the n-GaAs ,n-Al$_{0.1}$Ga$_{0.9}$As as well as n-Al$_x$Ga$_{1-x}$As doped to a Si concentration of 2.5×10$^{18}$ cm$^{-3}$. The mesa devices were processed using standard photo lithographic techniques. The top n-contact metal is formed by deposition of Pd/Ge/Ti/Pt/Au using electron beam evaporation. The bottom common electrode is p-contact metal made by Ti/Pt/Au.

The photocurrent spectra of the RP-LED are measured using the Fourier transform infrared spectrometer and the responsivities at different bias voltages shown in Fig. 1(c-d) are acquired using a calibrated blackbody together with a lock-in amplifierand a low noise current preamplifier. Owing to the ultrabroadband photoresponse in the RP, we measured the responsivities in two regions: 4-20 THz and 10.5- 225 THz, where Mylar films beam splitter and KBr beam splitter are used respectively. The optical window used in the cryostat for the two regions are high density polythene (HDPE: 4-20 THz) and TlBr-TlI (KRS-5: 10.5-225 THz) window, respectively.

The RP-LED shows an untrabroadband photoresponse in the frequency range of 4-200 THz, in which the NIR (90-200 THz) photoresponse is due to split-off band absorption (SOA) [25], MIR (15-90 THz) photoresponse is result from the free carrier absorption [26] and THz (4-15 THz) response originate from the hot hole injection induced cutoff frequency extending [27, 28, 29]. The photocurrent in Fig. 1(c-d) consists with forward photocurrent (I$_F$: consistent with the direction of the turn-on bias voltage of the LED) and ratchet photocurrent (I$_R$: opposite to the direction of the turn-on bias voltage of the LED) depending on the response mechanism in different frequency range and bias. The deep valleys at around 8 THz and 11 THz in Fig.1(c) corresponds to the GaAs optical phonon and AlAs-like phonon absorption. The other small valleys at high frequencies are due to the multiple-phonons absorption. The detail of the response characteristics of RP can be found elsewhere [27].

Fig.2(a) reveals that the applied bias voltage drops mainly across the LED part, and then the additional bias voltage goes to the RP part. When the applied bias is high enough to overcome the p-n junction in the LED and ratchet potential barrier in RP as shown in Fig.2(a), the holes could be able to response to the low frequency THz radiation [28, 29]. Since the hot hole related THz response can only be achieved after overcoming the built-in electric field in LED, all THz photocurrents appear as I$_F$ in Fig.1(c). The ratchet structure in RP-LED also results in a photovoltaic response characteristic. Under zero bias (in Fig.2(b)), IR photons stochastically change the momentum and energy of holes both in injector and absorber first. The absorption occurred here include the free carrier absorption (FCA) and split-off band absorption (SOA) [26, 25]. Then the asymmetry of the repeated unit of the barrier potential bias the motion of holes to the left due to the asymmetric relaxation [30]. This phenomenon is well known as the light-induced ratchet effect, which is regarded as a novel photovoltaic mechanism [30].

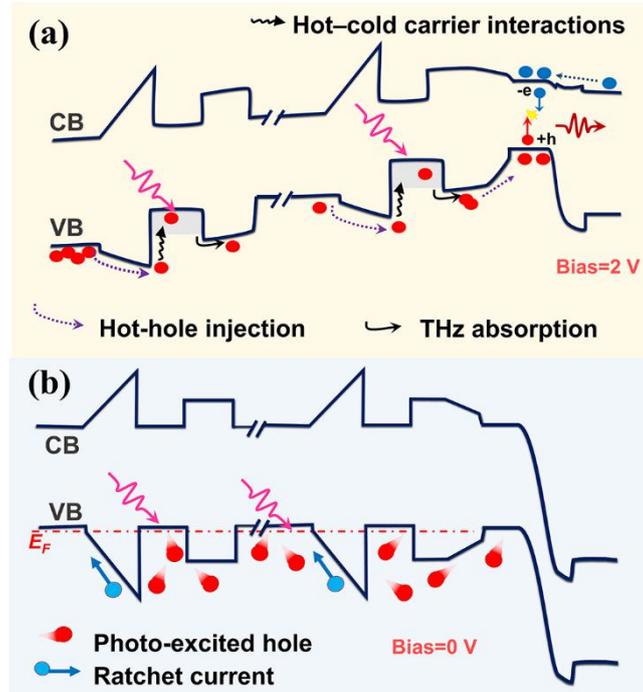

**Figure 2. (a) The band diagram and upconversion mechanism of RP-LED device with a forward bias higher than the turn on voltage of the device, (b) the band diagram and photovoltaic-like ratchet effect of RP-LED device at zero bias.**

We find that when the bias voltage is lower than 1V, the response of the RP-LED is basically the same as that at 0V (Fig.1(d)). This is because the photocurrent is mainly manifested as the ratchet photocurrent before the LED is turned on. When the LED is turned on and illuminated by IR/THz radiation, the resulting photocurrent drives the LED to emit NIR photons, which is collected from the top of the RP-LED. The inset of Fig.3 shows the emission spectrum of the LED at 4.2 K. The peak is at 837 nm, which is agree well with the bulk GaAs band gap taking the temperature and doping into consideration. The output electroluminescence (EL) power of the LED part is also displayed in Fig.3, which exhibits a turn-on behavior owing to the p-n junction in the LED structure.

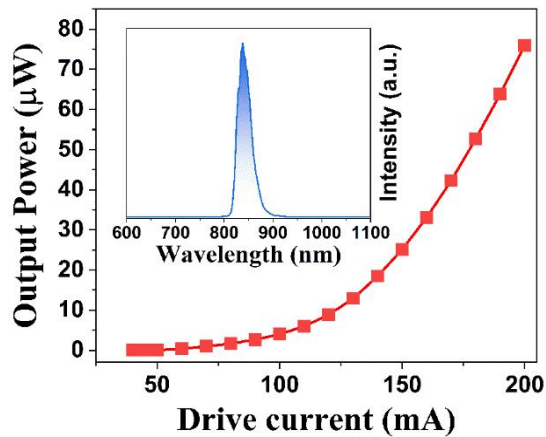

**Figure 3. Output electroluminescence (EL) power of the LED as a function of the drive current. The inset also displays the EL emission spectrum of the LED at 4.2 K measured by a fiber spectrometer (Ocean optics QE65PRO).**

The optical setup of the up-conversion experiment is presented in Fig.4(a). Radiation from the blackbody is collimated and focused by a pair of 90° off-axis parabolic mirrors. The reflected focal length of the off-axis parabolic mirror is about 4 inches. The RP-LED sample is placed at the focal position in the liquid helium cryostat. The front window of the cryostat is HDPE or KRS-5 to avoid the near-infrared or visible radiation influence. The rear window is quartz window to allow the transmission of the NIR photons. The emitted NIR power from the backside of the RP-LED is measured by the Thorlabs S130C large area Si slim photodiode. The power density of the blackbody source is shown in Fig.4(b), in which the two regions (4-20 THz 10.5-225 THz) are indicated using different colors.

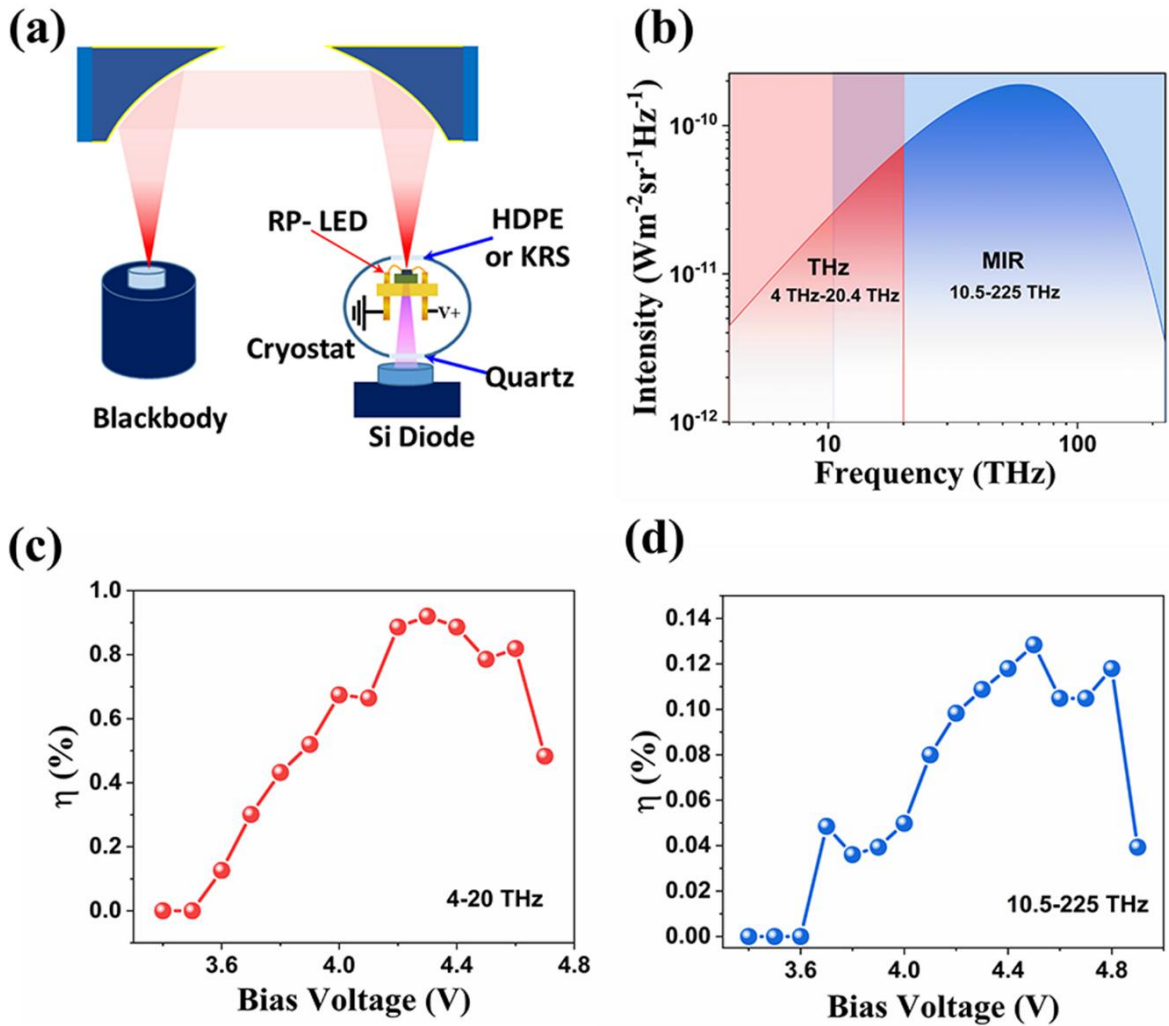

**Figure 4. (a) The optical setup of the up-conversion experiment, (b) the power density of the 1000 K blackbody source (Infrared Systems Development Corporation IR-564/301 with an emissivity higher than 0.99) and different color indicates the two transmission ranges of the optical window used in the cryostat, (c) the calculated upconversion efficiency ($\eta$) for the frequency range of 4-20 THz and (d) 10.5-225 THz.**

The broadband upconversion efficiency ($\eta$) is defined as the ratio of the number of output NIR photons to the number of incoming IR/THz photons:

$$\eta = \frac{N_{Emit}}{N_B} = \frac{P_{Emit}\lambda_{Emit}/hc}{\int_{v_{min}}^{v_{max}} \frac{S_M S_D 2hv^3}{L^2 c^2} \frac{1}{e^{hv/kT}-1} \frac{1}{hv} dv} \quad (1)$$

where $N_{Emit}$ is number of the output up-converted photons after the dark EL background is subtracted, $N_B$ is the number of the incoming IR/THz photons from the blackbody, $P_{Emit}$ is the measured NIR power from the LED part $\lambda_{Emit}$ is the wavelength of the NIR emission of the LED, $h$ is the Planck constant, $c$ is the velocity of light, $S_M$ is the collimating plane area of the 90° off-axis parabolic mirrors, $S_D$ is the detector area, $v$ is the frequency of the incoming IR/THz photon, $L$ is the reflected focal length of the off-axis parabolic mirrors, $k$ is the Boltzmann constant, $T$ is the temperature of the blackbody, $v_{min}$ and $v_{max}$ are determined by the frequency transmission range of the optical window of the cryostat (4-20 THz for HDPE window and 10.5-225 THz for KRS-5 window).

The upconversion efficiency ($\eta$) for the two frequency regions are shown in Fig.4(c-d). For both frequency range, the $\eta$ shows a turn-on behavior due to the p-n junction in LED part. The $\eta$ also shows an evident droop at high bias voltage, which is mainly caused by the LED efficiency droop due to the increased nonradiative recombination. The peak $\eta$ for the two frequency ranges is 0.92 % and 0.13 %, respectively. The $\eta$ is relatively low though the upconversion was successfully achieved. The upconversion efficiency is mainly suppressed by the LED, which is simply designed without any optimization. Future work can focus on the optimization of the barrier height, thickness and active layer doping and thickness of LED to achieve high-efficiency internal quantum efficiency. In addition, the introduction of InGaAs quantum wells in LED part can also reduce the internal reabsorption of the up-conversion device to further improve the internal quantum efficiency of the device. On these foundations, an appropriate light output coupling structure design for the LED can also be carried out to improve the light extraction efficiency of the LED so as to improve the up-conversion efficiency of the RP-LED device. Moreover, the quantum structure of the RP and the connection layer between the RP and LED also have plenty of improvement room for the high performance pixelless imaging application.

In conclusion, an ultrabroadband upconversion device is demonstrated, which is based on the integrated p-type GaAs/Al$_x$Ga$_{1-x}$As ratchet photodetector (RP) and GaAs double heterojunction LED (DH-LED). The normal incidence absorption simplifies the structure of the RP-LED device compared with the ISPT-based QWP-LED upconverter and make it more compact. An ultrabroadband photoresponse from THz to NIR region (4-200 THz) is realized that covers a much wider frequency range compared with the existing upconversion devices. Broadband IR/THz radiation from 1000 K blackbody is successfully upconverted into NIR photons. Additionally, the upconverter is also tested as photovoltaic detectors in the infrared region (15-200 THz) without an applied bias voltage due to the ratchet effect.

# Acknowledgement

This work was supported by the Natural Science Foundation of China (12104061, U1730246，

12074249，11834011 and 62004209), Natural Science Foundation of Shanghai (19ZR1427000), Project funded by China Postdoctoral Science Foundation (2020M680458), Open Project funded by Key Laboratory of Artificial Structures and Quantum Control (2020-03).

## Conflict of interest:

The authors have no conflicts to disclose.

## Data availability:

The data that support the findings of this study are available from the authors on reasonable request.

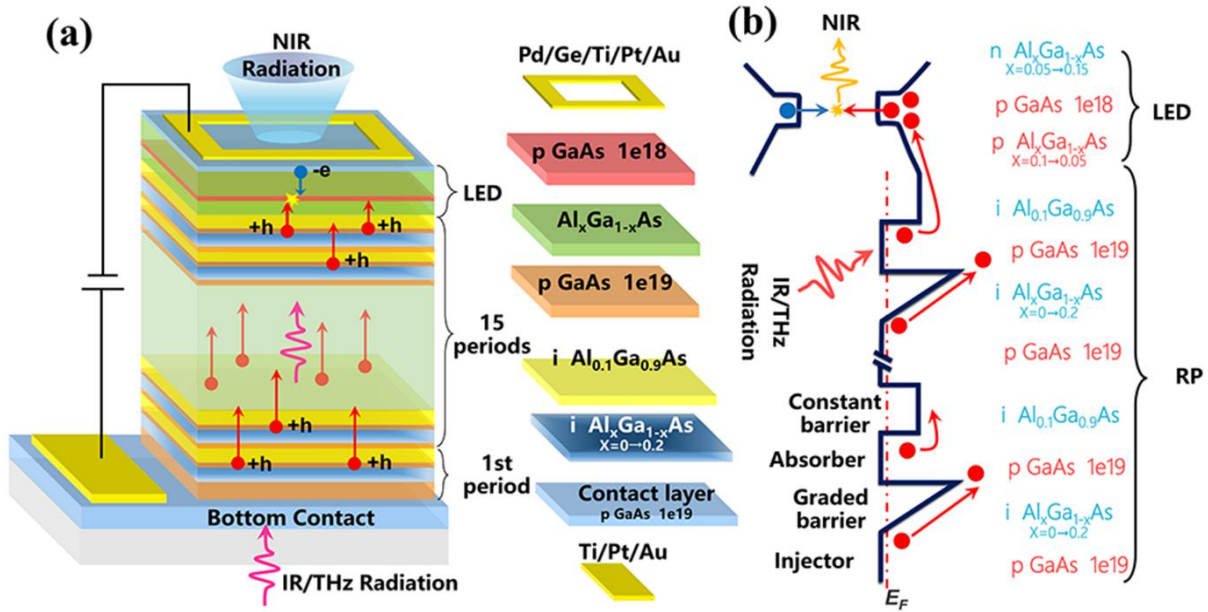
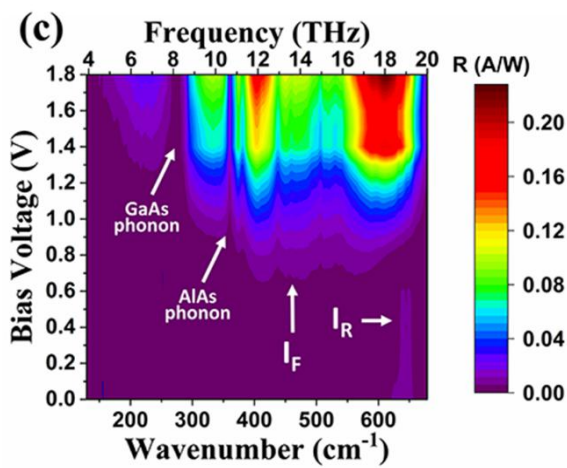
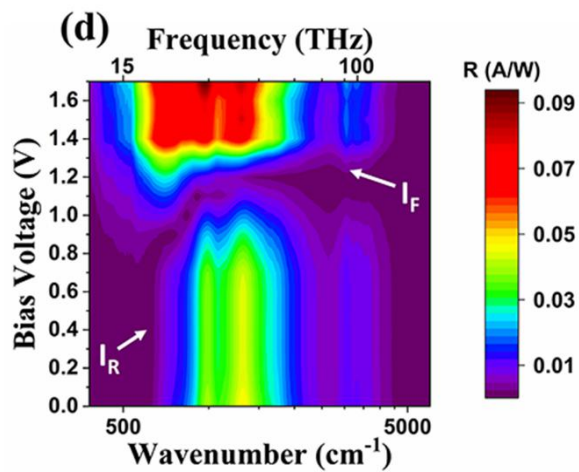

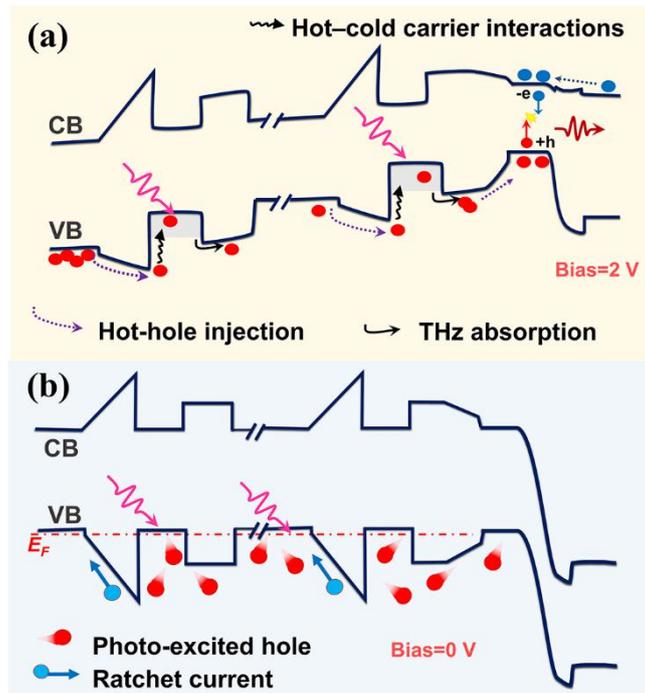

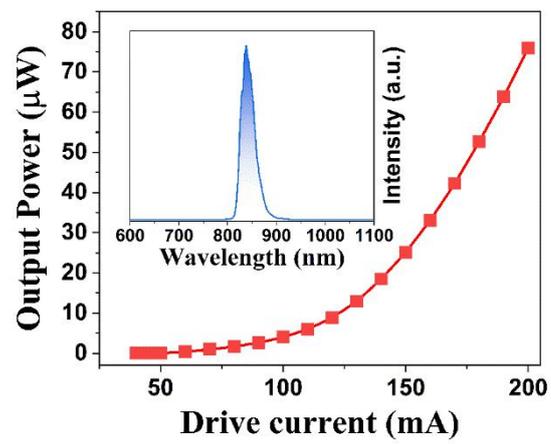

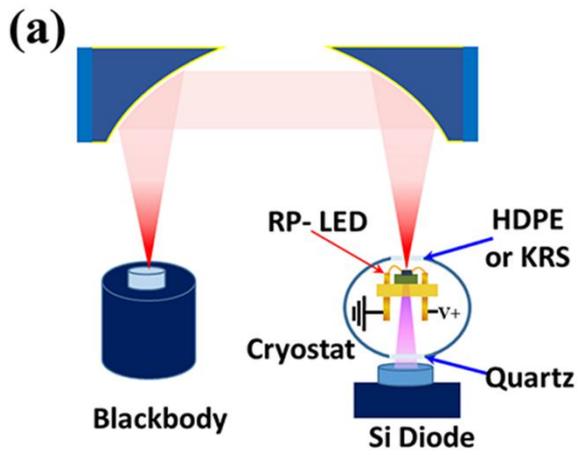
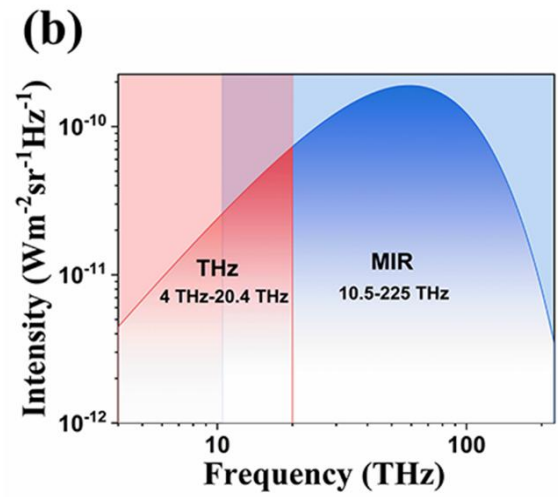
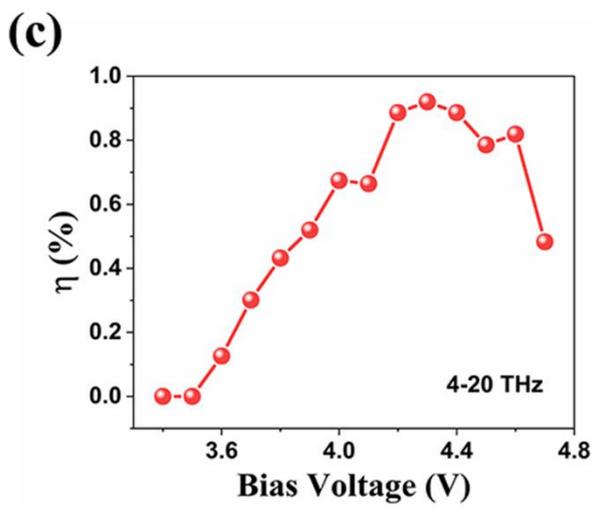
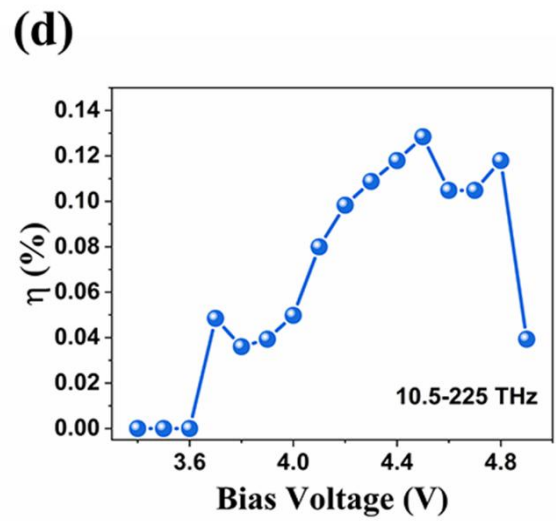